\documentclass[11pt]{article}
\usepackage{latexsym}

\def\be{\begin{equation}}
\def\ee{\end{equation}}
\def\bea{\begin{eqnarray}}
\def\eea{\end{eqnarray}}

\def\case#1/#2{\textstyle\frac{#1}{#2}}

\begin{document}

\vspace{.7in}

\begin{center}
\Large\bf {THE NATURE OF TIME}\normalfont

\vspace{.2in} \small(This article won the first prize awarded on
March 7th 2009 by the jury in the essay competition of the
Foundational Questions Institute (fqxi.org) on The Nature of Time.)
\\
\vspace{.2in} \normalsize \large{\textbf{Julian Barbour}}
\\
\normalsize \vspace{.1in} email: Julian.Barbour@physics.ox.ac.uk

\vspace{.2in}

\end{center}

\textbf{Abstract.} A review of some basic facts of classical
dynamics shows that time, or precisely \emph{duration}, is redundant
as a fundamental concept. Duration and the behaviour of clocks
emerge from a timeless law that governs change.

\section{Introduction}

My library contains four books on mechanics, the science of change
in time. Three of them -- all modern classics -- fail to define
either time or clocks! Relativity textbooks do discuss time and
clocks but concentrate on only one of the two fundamental problems
of time that Poincar\'e identified in 1898 \cite{Poincare}: the
definitions of \emph{duration} and of \emph{simultaneity at
spatially separated points}. Since then the first problem has been
remarkably neglected, probably because Einstein's solution of the
second in 1905 created such excitement.

The failure to discuss duration at a foundational level largely
explains the unease many feel when confronted with the idea that the
quantum universe is static. This suggestion emerged in 1967 from a
rather high-level attempt by Bryce DeWitt to meld Einstein's
\emph{classical} general theory of relativity \cite{Bryce} with
\emph{quantum} theory and has given rise to decades of agonizing
over `the problem of time'. In my view, had duration been properly
studied in classical physics, its disappearance in the conjectured
quantum universe would have appeared natural. In this essay I will
not discuss quantum theory at all but instead question the standard
assumptions made about duration in classical physics.

I shall develop from scratch a theory of time and clocks, linking it
to work that astronomers began in antiquity. The best guide to the
nature of time is the practice of astronomers. They cannot afford
mistakes; a missed eclipse is all too obvious. Moreover, they work
directly with concrete facts, their observations, not obscure
metaphysical notions of time.

My discussion begins with Newton's comments on astronomical practice
in his great \emph{Mathematical Principles of Natural Philosophy}
(1687), my only book on mechanics that does discuss duration.
Newton's discussion leads directly to two key intimately related
questions: How can we say that a second today is the same as a
second yesterday? What is a clock? The answers to these questions,
which are seldom addressed at a sufficiently foundational level,
will tell us much about time and the way the world works. We shall
find answers to them by examining successive important discoveries
made over two millennia.

I hope the answers to the two questions will persuade you that time
as an independent concept has no place in physics. It arises from
something concrete but deeper. As Ernst Mach said (1883)
\cite{Mach}:
\begin{quote}\small

It is utterly beyond our power to measure the changes of things by
time ... time is an abstraction at which we arrive by means of the
changes of things; made because we are not restricted to any one
definite measure, all being interconnected.
\end{quote}

Einstein, an admirer, quoted this passage in his obituary of Mach,
calling it a gem. Oddly, Einstein never directly attempted a Machian
theory of time, but in fact such a theory of `time without time'
sits hidden within the mathematics of his general theory of
relativity \cite{RWR}, the foundation of modern classical physics.

The time-without-time foundation of classical physics entails a
relatively small adjustment to our conceptions, but is likely to
have a profound effect in a quantum theory of the universe. This is
because significant parts of classical physics, above all time, are
carried over unchanged into quantum theory. If our ideas about time
in classical physics are wrong, surprises can be expected in a
quantum theory of the universe. As of now, little can be said about
this with certainty because no such theory yet exists. The very idea
could even be wrong. Nevertheless, candidate theories have been
proposed. The one I favour seems initially impossible: the quantum
universe is \emph{static}. Nothing happens; there is being but no
becoming. The flow of time and motion are illusions.

As I have said, I find this natural, but I shall not here describe
my vision of a timeless quantum universe presented in \emph{The End
of Time} \cite{EOT}. Instead, using a few elementary equations and
some new arguments, I wish to strengthen the case for eliminating
time as a fundamental concept in classical physics. The arguments
are simple. They strongly suggest that time should be banished.
Writing about the nature of time is a hard task. Unlike the Emperor
dressed in nothing, time is nothing dressed in clothes. I can only
describe the clothes.

\section{The Theory of Duration}

\subsection{Newton and the Equation of Time}

No essay on time can omit Newton's magisterial words:

\begin{quote} \small Absolute, true, and mathematical time, of itself,
and from its own nature, flows equably without relation to anything
external, and by another name is called duration: relative,
apparent, and common time, is some sensible and external measure of
duration by the means of motion, which is commonly used instead of
true time; such as an hour, a day, a month, a year.\footnote{For
readers finding difficulty with Motte's 1729 translation of Newton's
Latin, \emph{equably} means \emph{uniformly} and \emph{sensible}
means \emph{observable} (through the senses).}

\end{quote}
Newton here implies that in some given interval of true time the
universe could do infinitely many different things without in any
way changing that interval of time. This view still sits deep in the
psyche of theoretical physicists and only partly exorcized from
general relativity. I shall show that intervals of time do not
pre-exist but are \emph{created by what the universe does}. Indeed,
Newton can be hoist by his own petard if we see what his marvelous
laws actually tell us.

Let us start with his concession to practicality: the relative time,
which we are forced to use, is found ``by the means of motion''.
Moreover, the measures are concrete: shadows mark the hours on a
sundial; the moon waxes and wanes; the seasons pass. They are almost
as tangible as Shakespeare's ``daisies pied and violets blue'' that
come with the cuckoo. They are all clothes of time.

But what does Newton tell us about time itself, his ultimate
absolute along with space? Can you get your hands on time? He is
more aware of the question and a potential answer than many modern
authors:

\begin{quote} \small Absolute time, in astronomy, is distinguished
from relative, by the ... astronomical equation. The necessity of
this equation, for determining the times of a phenomenon, is evinced
as well from the experiments of the pendulum clock, as by eclipses
of the satellites of Jupiter.

\end{quote}

The `astronomical equation' is today called the \emph{equation of
time} and is the correction that equalizes -- hence `equation' --
the times measured by the sun or the stars. It is important for my
argument and so needs to be explained. The successive returns of the
sun and a given star to due south at the Greenwich Observatory
define the \emph{solar} and \emph{sidereal} days respectively. The
first is on average four minutes longer than the second because the
sun moves eastward relative to the stars along the great circle on
the sky called the \emph{ecliptic} (eclipses can only occur when the
moon too is on the ecliptic). Superimposed on the average
sidereal--solar difference are two effects. The sun's speed along
the ecliptic (measured by either day) is not exactly constant. This
reflects the earth's variable speed in its orbit around the sun.
Second, when the sun is high in the sky in the summer and low in the
winter, its motion along the ecliptic is purely eastward relative to
terrestrial south, and the sidereal--solar difference is enhanced
compared with the spring and fall, when this is not the case. The
two effects, well known in antiquity, cause solar time to be
sometimes ahead and sometimes behind sidereal time by about 15
minutes. Is there some reason to choose one of these times in
preference to the other?

When Ptolemy wrote the \emph{Almagest}, the compendium of ancient
astronomy, around 150 CE, he knew no modern laws of motion. In his
rudimentary astronomy the sun, moon, planets, and stars were all
carried around the earth in curiously different ways. Since the sun
outshone everything in the heavens, dominated life, and governed
civil order, he could have taken it to measure time. In fact, he
chose the stars; his reason is instructive and marks the first step
to a theory of duration.

He and his great predecessor Hipparchos (who flourished around 150
BCE) had developed a theory of the motion of the sun and the moon
around the earth. Its key element was uniform motion of both the sun
and moon around certain circles. It predicted eclipses of the moon
with reasonable accuracy provided the motion in the circles was
taken to be uniform \emph{relative to sidereal time}. Because 15
minutes are significant in eclipse prediction, the sidereal--solar
fluctuations ruled out the sun as the hand of time. What can we
learn from this? In his mind's eye, Ptolemy could `see' the stars,
sun, and moon moving in their circles. It is \emph{the way they
move}, specifically the \emph{correlations} between their motions,
that warrants the introduction of a distinguished measure of time.
Ptolemy's choice of sidereal time to measure duration remained
unchallenged for close on two millennia.

This is the justification of Newton's comments about the
astronomical equation. The earth's rotation was still by far the
best measure of time in his day, but astronomical knowledge had been
greatly extended, above all by Kepler's laws. The pendulum clock had
also been invented, and experiments confirmed that it -- and the
satellites of Jupiter -- marched better in step with sidereal than
solar time. In fact, Newton actually says that [absolute] duration
is to be distinguished ``from what are only sensible measures
thereof ... by means of the astronomical equation''. This statement
is both important and ironic; important in confirming that the
measure of time is not chosen before but after the discovery of
specially correlated motions, ironic because the motion of the stars
is just as `sensible a measure' as any other. As Newton himself
defines it, absolute time is by no means independent of the world;
it is a specific motion, the rotation of the earth.

\subsection{The Changes of Things, All Interconnected}

Two comments before we proceed. Since time must be deduced from
change of position (motion), I shall here take position and
differences of position as given, though great issues do lurk behind
these apparently simple notions \cite{DOD}. Second, clocks are of
two kinds: natural like the earth's rotation and man made. Because
man-made clocks are complex and rely on special devices, they do not
reveal the nature of time and the basis of time keeping as well as
natural clocks. For example, we shall see that the often made
statement that a periodic process is the basis of a clock is
misleading.

The development of astronomy in the period from Newton to the end of
the 19th century nicely illustrates Mach's words recalled above. For
the sake of a telling image, let us `simplify' the solar system and
suppose all the planets revolve around the sun in one plane; that
the earth's rotation axis is perpendicular to that plane; and that
astronomers observe them against the background of the stars from a
`crow's nest' very far `above' the sun. From it they can observe
directly: the distances $r_i$ of each planet $i$ from the sun; the
angle $\phi$ through which the rotating earth turns relative to a
fixed star ($\phi$ measures terrestrial sidereal time); the angles
$\alpha_i,~i=1, 2, ...,$ between the lines from the sun to each
planet $i$ and a fixed star. Modern textbooks, leaving us us to
fathom the meaning of $t$, say that all these quantities are
\emph{functions of the time}: $\phi(t),\, \alpha_i(t),\, r_i(t)$.
But Newton effectively identified time with sidereal time, i.e., the
angle $\phi$. In reality the Newtonian specification of each
observed configuration of the solar system is
$\alpha_i(\phi),\,r_i(\phi)$, i.e, certain values of $\alpha_i$ and
$r_i$ for each value of $\phi$. The undefined $t$ plays no role.

Now we have to ask why the earth's rotation angle plays such a
distinguished role. The earth can hardly be the lord of the planets'
dance. Why not some other motion? Two great discoveries cast light
on this question and did for a while bring in other special
candidate motions.

On Easter Sunday 1604 Kepler's immense labour finally crystallized
in his first two laws of planetary motion : the planets move in
ellipses with the sun at one focus; the line from the sun to each
planet sweeps out equal areas in equal times \emph{as measured by}
$\phi$. We now have something special. The motions of the planets
could have been arbitrary, but no: they exhibit remarkable
\emph{correlations}. The areas swept out by the planets and the
earth's rotation all march in step. They `keep the same time'.
Moreover, the correlations are only found between certain motions;
there are no clean correlations between the varying planet--sun
distances $r_i$.

The second discovery, unthinkable without Kepler, was Newton's laws
of motion and universal gravitation. His first theorem in the
\emph{Principles} uses Kepler's area law to prove that all the
planets are deflected from their natural inertial motion by a force
that tugs them toward the sun. And what measure now does Newton take
as time to find the acceleration, the second time derivative,
generated by this centripetal force? In one truly beautiful
geometrical proof, he takes advantage of the further specially
correlated motions found by Kepler and uses, not the earth's $\phi$,
but the area swept out by the very planet whose acceleration he is
determining. Newton had discovered dynamics. Wonderfully simple laws
governed all motions in the heavens and on the earth.

But had he caught time? Clearly no, but still a great catch: diverse
and special precisely correlated motions and a law of gravitation
that gave the observed accelerations relative to these motions.
Newton was wrong to go beyond this fact, tempting though it was to
`see' the invisible structure of time behind its clothes.

Mach in contrast was right: we do abstract time from motion. It only
seems to be a universal absolute ``because we are not restricted to
any one definite measure, all being interconnected''. For Newton's
purposes, the area swept out by a planet was as good as $\phi$. But
then it turned out that this is not strictly true. In accordance
with Newton's laws and as confirmed observationally in the 18th and
19th centuries, the planets perturb each other's motions. Their
orbital areas do not march \emph{perfectly} in step with $\phi$ or
with each other. Remarkably, using $\phi$, the `sensible' hand that
Newton had ironically identified as absolute time, the calculated
perturbations matched the observations perfectly. The astronomers
were lucky; no man-made clock could remotely rival the earth's
rotation at that time.

This happy situation persisted until the 1890s, when a crisis
developed whose resolution takes us deeper into the nature of time.

\subsection{The Acceleration of the Moon and Ephemeris Time}

The moon moves across the stellar sky faster than the other
celestial bodies and is much easier to observe and use to test
Newton's laws. In the 1890s, astronomers came to the uncomfortable
conclusion that the moon exhibited a small but undeniable
non-Newtonian acceleration. What could be the cause? They wondered
whether the earth might absorb the sun's gravity during eclipses of
the moon and allow the anomalous acceleration, or whether the moon's
tidal effects could be slowing the earth's rotation.

Most astronomers correctly guessed the latter but then had to seek a
better and more fundamental measure of time. I shall argue that,
perhaps without realizing it, they implicitly asked the ultimate
question: \emph{what is time?} As Poincar\'e described it in his
important paper \cite{Poincare}, they proceeded as follows. Suppose
Newton's laws are correct and the solar system is a closed dynamical
system, i.e., no external objects exert significant disturbing
forces on it. Then it must be possible to define a time variable
such that Newton's laws do hold for the solar system when it is
used. The astronomers \emph{defined} time so as to ensure that the
laws hold \cite{Clemence}.

There is a short cut that enables us to `see' this time. It exploits
a fundamental concept: the energy of a system. Here we need the
equations. Mutually gravitating bodies have potential energy

\begin{equation}
V=-G\sum_{i<j}{m_im_j\over r_{ij}},\label{V}
\end{equation}
where $G$ is Newton's gravitational constant, $m_i$ is the mass of
body $i$ in the system, and $r_{ij}$ is the distance between bodies
$i$ and $j$. The summation symbol $\sum$ with $i<j$ under it means
take all pairs $ij$ once, calculate $m_im_j/r_{ij}$ for each, and
add them all together.

The bodies also have kinetic energy $T$, which is simply the sum of
the kinetic energies of each body (half its mass $m_i$ times the
square of its speed $v_i$). I want to write this in the most
revealing way. The Greek letter $\delta$ before some symbol, say $d$
for distance, means $\delta d$ is very small. Now suppose that, in
the small time $\delta t$ to be defined, body $i$ moves the distance
$\delta d_i$. Then to a good approximation its speed is
$$
v_i={\delta d_i\over\delta t},
$$
and
$$
T=\sum_i{m_i\over 2}\left({\delta d_i\over\delta t}\right)^2,
$$
where now the contributions of each particle are added.

Perhaps the most fundamental result in dynamics is that the energy
$T+V$ of an isolated system remains exactly equal to a constant $E$
in time. One says that the energy is conserved. This is normally
expressed by the equation
\begin{equation}
\sum_i{m_i\over 2}\left({\delta d_i\over\delta
t}\right)^2+V=E,\label{C}
\end{equation}
where $V$ is given by (\ref{V}).

Suppose the astronomers in the crow's nest take `snapshots' of the
solar system in successive configurations. Each gives directly the
separations $r_{ij}$, and from two taken in quick succession the
$\delta d_i$ can be found. If the astronomers had a clock with them,
they could test the law (\ref{C}). But in the 1890s they had no
adequate clock. What they effectively did was rearrange (\ref{C})
and use it to \emph{define} the increment $\delta t$ of time between
successive configurations:
\begin{equation}
\delta t=\sqrt{\sum_im_i(\delta d_i)^2\over 2(E-V)}.\label{TIME}
\end{equation}

The time derived from (\ref{TIME}) is called \emph{ephemeris time}
\cite{Clemence}; an ephemeris gives the positions of celestial
bodies, and $\delta t$ is deduced from such positions. We are now in
a position to define a man-made clock, a \emph{chronometer}: ``it is
a mechanism for measuring time that is continually synchronized as
nearly as may be with ephemeris time'' \cite{Clemence}. The
astronomer Clemence was led to propose this definition in 1957
because he found that many authors writing about relativity have
``no clear idea of what a clock is.''

Apart from the constants $G$ and $E$ and $m_i$, which I shall
discuss shortly, (\ref{TIME}) contains only displacements and
distances. According to it, the instantaneous speed of particle $i$
is
\begin{equation}
v_i={\delta d_i\over \delta t}=\sqrt{2(E-V)\over\sum_im_i(\delta
d_i)^2}~\delta d_i.\label{SPEED}
\end{equation}

The $\delta t$ appearing here in $\delta d_i/\delta t$ is merely
shorthand for (\ref{TIME}), and the expression on the right is the
one that counts. The elusive $t$ has been eliminated. Science should
deal with observable things, so this is a step forward. Equation
(\ref{SPEED}) expresses the truth that only relative quantities have
objective meaning. Speed of body $i$ is not the ratio of its
displacement to an abstract time increment but to (\ref{TIME}),
which involves the displacements of all the bodies in the system.

I regard the definition of duration by (\ref{TIME}) as exceptionally
important for two reasons: it is made possible solely because the
bodies in the system move in a very special way (duration does not
`pre-exist' that fact as Newton asserted); time is no longer
measured by particular individual motions but by the sum of all
motions.

There is however still some work to do to get that time concretely.
If the astronomers know the values of $G$, $E$, and $m_i$ in
advance, two `snapshots' taken of the solar system in quick
succession will suffice to determine the $\delta t$ between them. If
they do not, a sufficient number must be taken to provide enough
data for their determination. The `time' (\ref{TIME}) will then
truly emerge from observed positions of objects. Time can be read
off the heavens.

Valuable as is the astronomers' notion of ephemeris time, the
principles behind it are still somewhat vague. The astronomers start
with a theory formulated in terms of an undefined time and then use
observations to define it. I want to show in the remainder of the
paper that this obscures the truth. It is evident from what I have
so far presented that one could never speak about a time worthy of
the name were it not for the wonderfully correlated motions that
nature exhibits. What we really need is a \emph{timeless} theory of
the correlations. Before providing that I want to make clear just
how remarkable the correlations are. This will have the advantage of
highlighting the difference between Einstein's definition of a clock
and Clemence's \cite{Clemence}.

\section{Contrasted Definitions of Clocks}

An important question that we have not yet considered is how
\emph{natural} clocks can march in step. By Clemence's definition,
man-made chronometers will do that through human artifice since they
must all be synchronized to the solar-system ephemeris
time.\footnote{In fact, it is now atomic time. Astronomers replaced
sidereal time by ephemeris time only in 1952 and then introduced
atomic time in 1979.} A point that I want to make now is that, from
the fundamental perspective, it is a mistake to concentrate on the
definition of a single clock. Clocks are useless if they do not
march in step for otherwise we cannot keep appointments. Therefore
it is not \emph{a} clock that we must define but \emph{clocks and
the correlations between them} as expressed in the marching-in-step
criterion. We shall see that this leads to a very different
definition of clocks -- and understanding of time -- to that of
Einstein, given below.

A crucial role in the definition of natural clocks is played by the
term $E-V$ in (\ref{TIME}). To my knowledge, its significance,
through rather obvious, has not hitherto been noted. We imagined our
astronomers `looking down' on the solar system. Suppose that far
`above them' is another system, similar but not identical. Its
energy $E$ and the masses of its bodies will be different. Suppose
the astronomers take simultaneous snapshots of the two systems at
successive instants and want to determine the time interval between
the instants from the snapshots. Now in general the value they
obtain of $\delta t_2$ will not be equal to $\delta t_1$ because the
quantities that go into (\ref{TIME}) for the two systems can be very
different. However, the two systems will be good clocks if in any
interval the ratio $\delta t_2/\delta t_1$ is always the same. For
then they march in step but merely take larger steps like the tips
of ticking second hands of unequal lengths.

It is here that $E-V$ comes into play. In any dynamical system,
including the gravitational ones we are considering, the kinetic
energy $T$ and potential energy $V$ are constantly changing, but
their sum $E=T+V$ remains constant -- the total energy is conserved.
As a result, if $T$ gets larger, so does $E-V$. This is now crucial
for the way our two `clocks' behave. If $T$ for system 1 becomes
relatively larger than $T$ for system 2, the numerator in
(\ref{TIME}) of system 1 will be boosted compared with system 2 but
the denominators will counteract the effect; the ratio $\delta
t_2/\delta t_1$ will always be same, and the two `clocks' will keep
step. This will be true for any number of systems.

Thus, \emph{natural clocks are isolated dynamical systems with hands
that are advanced in accordance with} (\ref{TIME}); the law that
governs their behaviour -- which I have still to spell out in
timeless form -- ensures the marching-in-step condition.

Finally, in reality there is no \emph{perfectly} isolated system
except the entire universe. In a Newtonian picture, it could be the
multitudinous stars of our Galaxy in otherwise infinite space. Of
course, in such a universe there can be innumerable effectively
isolated systems and as many synchronous ephemeris times.

We see from this that time has no role to play as an independent
element of reality. It can be abstracted from change. However, Mach
could have been more specific -- the definition of time must be
based on an accurate description of the way naturally occurring
motions are correlated. Ptolemy's successful epicycle theory
selected sidereal time in preference to solar time; equally Newton's
success with his dynamical laws justified his claim that `absolute
time' could be deduced from the astronomical equation. When we push
our understanding to the limit, we arrive at the ephemeris time
(\ref{TIME}) with its reduction of time to masses, displacements,
separations, and the constants $E$ and $G$. Even in Einstein's much
more sophisticated general relativity time emerges in much the same
way \cite{RWR}.

The theory of duration and clocks that emerges from observable
differences as made explicit in (\ref{TIME}) is very different from
the view that prevailed among the great relativists in the early
20th century. It will suffice to consider Einstein's definition of
clock in 1910 \cite{Einclock}:
\begin{quote}

\small By a clock we understand anything characterized by a
phenomenon passing periodically through identical phases so that we
must assume, by the principle of sufficient reason, that all that
happens in a given period is identical with all that happens in an
arbitrary period.

\end{quote}

I see several problems with this definition. First no system ever
runs through truly identical phases, so this is an idealization that
hides the true nature of time; it lacks Poincar\'e's insistence
\cite{Poincare}, which I have repeated, that only the universe and
all that happens in it can tell perfect time. Second, Einstein's
clock cannot measure time continuously. It can only indicate that a
given interval has elapsed when identical phases recur. It can say
nothing about the passage of time in intervals within phases. Since
the universe is the only perfect clock, it seems nothing at all can
be said about the passage of time unless there is recurrence of eons
under identical conditions. Even then one can only say that the eons
are equally long. In contrast, the ephemeris time defined by
(\ref{TIME}) runs continuously and in no way relies on recurrence of
identical phases. Finally, by relying on periodicity, Einstein's
definition fails to identify the true dynamical basis of time
keeping and the importance of understanding why clocks can march in
step.

Now to the timeless law that explains how billions upon billions of
natural clocks scattered through the vast reaches of space can all
tick in step.

\section{The Timeless Principle of Least Action}

\emph{The configuration space of the universe} is the key concept.
In a universe of $N$ particles, each particle position has 3
coordinates; $3N$ define a complete configuration, which corresponds
to a unique \emph{representative point} $p$ in an abstract
$3N$-dimensional space $\cal U$. As the universe's configuration
changes, $p$ traces a curve in $\cal U$. If the universe were to
follow some arbitrary curve, no law would hold. The remarkably
simple and beautiful \emph{principle of least action} singles out
the special curves in $\cal U$ for which Newton's laws hold. I shall
present it in a little-known timeless form: \emph{Jacobi's
principle}.

You choose in $\cal U$ two points -- two configurations of the
universe. These are to remain fixed. You consider all possible
\emph{trial curves} that join them continuously in $\cal U$. In the
usual formulation, all trial curves are assumed to be traversed in a
fixed pre-existing time interval. No such assumption is made here;
it is unnecessary. All you need to do is divide each trial curve
into very short segments. For each segment, you calculate its
\emph{action}
\begin{equation}
\delta A=\sqrt 2\sqrt{\left(E-\sum_{i<j}{m_im_j\over
r_{ij}}\right)\sum_im_i(\delta d_i)^2},\label{AS}
\end{equation}
where $\delta d_i$ is the distance particle $i$ has moved. Here $E$
is not regarded as an energy but as a fundamental constant (like the
cosmological constant $\Lambda$ \cite{RWR}). The action $A$ for the
complete trial curve is the sum of the actions $\delta A$ for each
segment; in the limit in which the segments are made shorter and
shorter, $A$ tends to a finite limit.

Now comes the wonderful thing. For one of the trial curves, the
action will be smaller than for any other. For this \emph{extremal}
curve, and in general for no other joining the fixed end points,
\emph{the particles obey Newton's laws with the emergent time
defined by} (\ref{TIME}). This is a timeless law; it determines a
path, or history, in $\cal U$. The key thing is that no time is
assumed in advance. A time worthy of the name does not exist on any
of the non-extremal curves. Time emerges only on the extremal
curves.

It is not only Newton's laws that can be obtained in this timeless
way. There is an interpretation of Einstein's general relativity in
which it and time arise in much the same way \cite{RWR}. I will not
claim that time can definitely be banished from physics; the
universe may be infinite, and black holes present some problems for
the timeless picture. Nevertheless, I think it is entirely possible
-- indeed likely -- that time as such plays no role in the universe.

Occam's razor tells us to avoid redundant elements. All we need are
differences. Indeed, the passage of time is always marked by
difference, often seen as cruel:

\begin{quote} \small
When forty winters shall besiege thy brow,\newline And dig deep
trenches in thy beauty's field,\newline Thy youth's proud livery, so
gazed on now,\newline Will be a tattered weed of small worth held.

\end{quote}

Unlike Newton, Shakespeare did not attempt to describe time itself,
only the differences associated with it. Look again at the
`expression for time':
$$
\delta t=\sqrt{\sum_im_i(\delta d_i)^2\over 2(E-V)}.
$$
Some of those $\delta d_i$s are the trenches dug in youth's brow.

\vspace{.2in}

\small \textbf{Acknowledgements.} My thanks to Nancy McGough, Naomi
Barbour and especially Boris Barbour for critical comments that
improved this essay. I am grateful to the jury for awarding me the
prize for this essay and to the Foundational Questions Institute
(fqxi.org) for organizing the essay competition and for support of
my research through grant RFP2-08-05.

\end{document}